\documentstyle[epsfig]{l-aa}
\input epsf
\newcommand{\nh}{${\mathrm N_H}$}
\newcommand{\cgs}{{\mathrm ~erg~cm^{-2}~s}^{-1}}

\newcommand{\chandra}{{\it Chandra} }
\newcommand{\rosat}{{\it ROSAT} }
\newcommand{\asca}{{\it ASCA} }
\newcommand{\sax}{{\it Beppo}SAX }

\newcommand{\lsim}{{\lower.5ex\hbox{$\; \buildrel < \over \sim \;$}}}
\newcommand{\gsim}{{\lower.5ex\hbox{$\; \buildrel > \over \sim \;$}}}

\begin{document}

   \thesaurus{         
              (; 
               )}
   \title{Extragalactic 2-10 keV source counts from a fluctuation 
analysis of deep BeppoSAX MECS images}

%   \subtitle{}

   \author{M. Perri\inst{1,2}
   \and P. Giommi\inst{1}}

   \offprints{M. Perri}

\institute{
   {\inst{1}\sax Science Data Center, ASI,
    Via Corcolle, 19,
    I-00131 Roma , Italy}\\
   {\inst{2}Dipartimento di Fisica, Universit\`a di Roma ``Tor Vergata'',
    Via della Ricerca Scientifica, 1,
    I-00133 Roma, Italy}\\
} 

   \date{Received  ; accepted }

   \maketitle
   \markboth{M. Perri and P. Giommi: 2-10 keV counts from a 
fluctuation analysis of deep \sax MECS images}{}

\begin{abstract} 

We present an analysis of the spatial fluctuations of the 2-10 keV 
Cosmic X-ray Background (CXB) as measured from 22 high galactic 
latitude ($|{\rm b}| > 25\degr$) fields observed with the MECS instrument 
on-board {\it Beppo}SAX. This technique allowed us to probe 
extragalactic source counts a factor 3-4 fainter than is possible 
with direct measurements of pointlike sources in MECS deep fields.
The slope of the 2-10 keV log$N$-log$S$ relationship is found to be still 
close to the ``Euclidean'' one ($\gamma = 1.5$) down to our flux limit of 
$\sim 1.5\times10^{-14} \cgs$, where the contribution of 
discrete sources to the 2-10 keV CXB amounts to $\sim 40-50\%$. 
Source counts derived from the analysis presented in this letter are in 
very good agreement both with those directly measured with \asca 
and \sax deep surveys at bright fluxes and with a first estimation of 
the 2-10 keV \chandra log$N$-log$S$ at fainter fluxes. 
\end{abstract} 

 
\section{Introduction}

The understanding of the origin of the Cosmic X-ray Background 
(CXB, Giacconi et al. 1962) has been and still is one of the main 
topics of X-ray astronomy. It is now widely accepted that the CXB is 
the results of the superposition of the emission of faint extragalactic 
discrete sources and many surveys of the X-ray sky have been carried 
out  to directly resolve the CXB and study the nature of these sources.

The soft X-ray energy band (0.5-2 keV) has been extensively 
investigated with \rosat with which a large fraction of the soft CXB 
($70-80\%$) has been resolved into discrete sources (Hasinger 
et al. 1993, 1998). 
Spectroscopic identifications of their optical counterparts have 
revealed that the large majority ($>80\%$) are AGNs, mostly with 
broad emission lines (Schmidt et al. 1998). 

The harder 2-10 keV sky has become accessible to deep surveys 
only over the past few years thanks to the imaging instruments of  
\asca and {\it Beppo}SAX. About 30\% of the CXB in this band has 
been directly resolved into discrete sources and the log$N$-log$S$ 
relationship has been derived down to a flux limit of $\sim 5 \times 
10^{-14} \cgs$ (Ueda et al. 1998; Cagnoni, Della Ceca \& Maccacaro 
1998; Giommi et al. 1998; Della Ceca et al. 1999; Ueda et al. 1999a, 
1999b; Fiore et al. 1999; Giommi, Fiore \& Perri 1999; 
Fiore et al. 2000a; Giommi, Perri \& Fiore 2000). 

A number of deeper surveys carried out with \chandra are now 
starting to probe the 2-10 keV CXB at much fainter fluxes. However, 
at least for some time, 
these can only cover a small area of sky with the consequence 
that the statistics on source counts (and optical identifications) is 
still very limited (Brandt et al. 2000; Fiore et al. 2000b; 
Mushotzky et al. 2000; Hornschemeier et al. 2000).

In this letter we present the results of a fluctuation analysis of \sax 
MECS images performed in the 2-10 keV energy band that allowed us 
to study the log$N$-log$S$ down to fluxes of $\sim 1.5\times10^{-14} \cgs$. 
The analysis of the spatial fluctuations of the CXB is a powerful 
method that allows us to investigate the log$N$-log$S$ relationship 
beyond the flux limit where single sources became too faint, or confused, 
to be directly detected. 
The theory behind this technique has been described for the first time 
by radio astronomers (Scheuer 1957, 1974; Condon 1974) and many 
applications of this method to X-ray data have successfully predicted 
source counts at faint fluxes (Hamilton \& Helfand 1987; Barcons \& 
Fabian 1990; Hasinger et al. 1993; Barcons et al. 1994; Butcher 
et al. 1997; Gendreau, Barcons \& Fabian 1998). 
Basically, the method consists of i) comparing the observed spatial 
fluctuations distribution of the CXB with a set of model distribution 
curves corresponding to different hypotheses on the power law slope 
$\gamma$ and normalization $K$ of the log$N$-log$S$ relationship and ii) 
by means of a fitting procedure deriving best fit parameters on source 
counts. 

The main contribution to the observed CXB spatial fluctuations comes 
from sources at a flux level $S$ corresponding to a source density 
of about 1-2 sources per beam (Scheuer 1974). This condition determines 
the sensitivity limit of this technique. 

\section{Data analysis}

The instruments aboard the \sax Satellite (Boella et al. 1997a) include 
a Low Energy (0.1-10 keV) Concentrator Spectrometer 
(LECS, Parmar et al. 1997), a Medium Energy (1.3-10 keV) Concentrator 
Spectrometer composed by three units (MECS, Boella et al. 1997b), a 
High Pressure Gas Scintillation Proportional Counter (3-120 keV) 
(HPGSPC, Manzo et al. 1997) and the Phoswich Detector System (15-300 keV) 
(PDS, Frontera et al. 1997). All these instruments point in 
the same direction and are collectively called Narrow Field Instruments 
(NFIs). In addition \sax carries two wide field (20 $\times$ 20 degrees 
FWHM) instruments operating in the 2-30 keV energy band (Wide 
Field Cameras, WFC, Jager et al. 1997) and pointing in diametrically 
opposed directions perpendicular to the NFIs. 

The data used in our fluctuation analysis consist of 22 
non-overlapping high galactic latitude ($|{\rm b}| > 25\degr$) MECS fields 
pointed at ``blank'' parts of the sky, that is not centered on 
previously known X-ray targets. About half (13) of these fields 
are Secondary Pointing NFIs observations, produced when the primary target 
(usually the Galactic Center) was observed with one of the Wide Field Cameras. 
The other observations include 3 non-overlapping fields centered 
near Polaris, the \sax default pointing position in case of safe mode, 
2 follow-up observations of Gamma Ray Bursts (GRBs) carried out 
early in the mission in which X-ray afterglows of the GRBs were not 
detected, 3 images, taken from the public archive, not centered on 
known X-ray sources, and 1 image taken during the Leonids 
meteorites crossing on November 17, 1999,  when the satellite was 
oriented so as to minimize the probability of impact with one of the 
meteorites. 
\begin{figure} 
\epsfig{figure=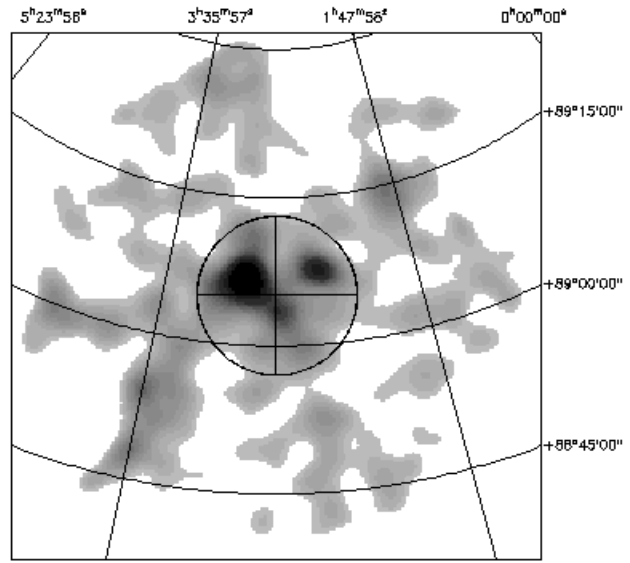,height=8.7cm,width=8cm,angle=-90} 
\caption{MECS 2-10 keV image of one of the ``blank'' fields used 
in our analysis. The 4 equal quadrants in which the central 8 
arcminutes circular region has been divided are also shown.}
\label{mecs_fov} 
\end{figure} 
The average exposure of the 22 fields is $\sim 88~{\mathrm ks}$ while 
typical values of galactic \nh, as derived by the 21 cm measurements of 
Dickey \& Lockman (1990), are $\sim 4 \times 10^{20}~{\mathrm cm}^{-2}$. 
No point sources with 2-10 keV flux greater than $5 \times 10^{-13} \cgs$ 
are detected in any of these fields.

As discussed above, the MECS instruments is composed of three nearly 
identical units. One of these (MECS1) failed on May 7th 1997. To 
ensure uniformity we have used data only from the units MECS2 and 
MECS3, to which we refer hereafter as the MECS instrument. To avoid 
difficulties due to complex variations in sensitivity 
across the field-of-view (FOV) we have restricted our analysis to 
the central 8 arcminutes circular regions (see Figure 
\ref{mecs_fov}) where the sensitivity is higher and vignetting and 
Point Spread Function (PSF) variations are small. 

In order to enhance the sensitivity of our analysis as well the 
statistics of the CXB granularity each circular region 
has been divided in 4 equal quadrants (each covering an area of 
$\sim 50.3~{\mathrm arcmin}^2$, see Figure \ref{mecs_fov}) for a total of 
88 independent measurements of the CXB. This sets our sensitivity to 
$S\sim 1.5 \times 10^{-14} \cgs$, the exact value depending on the 
details of the log$N$-log$S$ relationship. At this flux level 
one source corresponds to $\sim 10-15$ MECS net counts, well above the 
``1 photon per source limit'' which is considered as a further limit 
of the fluctuation analysis method.

\subsection{The Non X-ray Background}

To estimate the amount of cosmic X-rays we must subtract the 
non X-ray background (NXB, that is the instrument noise plus the 
particle background) which, in the central 8 arcminutes of the MECS, 
accounts for about half of the total counts. 

\begin{figure} 
\epsfig{figure=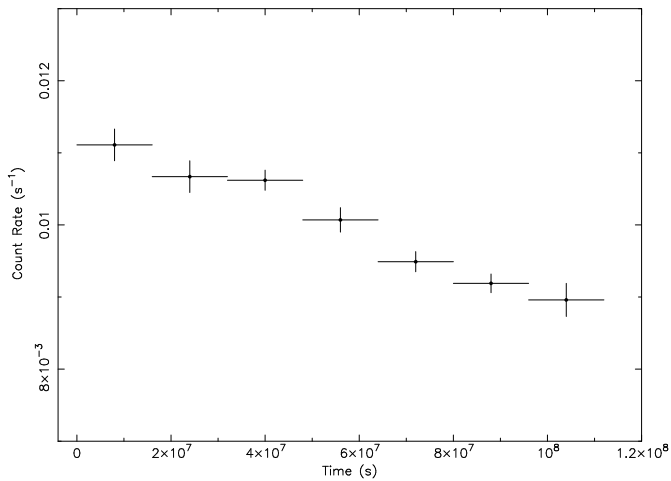,height=9.7cm,width=7.2cm,angle=-90} 
\caption{MECS Non X-ray Background count rate (channels 44-200, 
central 8 arcminutes circular regions) as a function 
of time. ``Dark Earth'' data have been accumulated from 1996 July 6 
($t=0$) to 1999 September 11 for a total exposure time of 3.85 Ms.} 
\label{mecs23_lc} 
\end{figure} 

To measure the NXB we have used, as in Parmar et al. (1999) and in
Vecchi et al. (1999), \sax MECS ``dark Earth fields'', that is MECS 
images taken  when the pointing direction is occulted by the part of 
the Earth that is not illuminated by the Sun. 
We have accumulated ``dark Earth'' observations from July 1996 through 
September 1999 for a total exposure time of 3.85 Ms. Counts between 
channels 44 and 200 (the energy range used to estimate the CXB, see 
below) have been extracted from the inner 8 arcminutes circular regions 
of the images. The corresponding MECS light curve is shown in Figure 
\ref{mecs23_lc}. A gradual decline of the count rate with time is evident. 
From July 1996 to September 1999 the amount of this reduction is 
$\sim 15-20\%$. We modeled this behaviour with the law 
$cr(t) = cr(0) - 2.07 \times 10^{-11} \times t$, where $t$ is time in seconds 
and $cr(0) \simeq 1.11 \times 10^{-2}~{\mathrm s}^{-1}$ is the NXB count rate 
value at the beginning of the mission ($t=0$, 1996 July 6). 
A similar instrumental background reduction has been reported for 
the LECS instrument on-board \sax by Parmar et al. (1999).

We also investigated potential systematic errors in the estimation of the 
NXB, such as significant deviations of count rate from the law given above 
or the statistical poisson noise of the NXB counts in a MECS quadrant. In 
both cases, given the typical exposure times of our MECS fields 
($T \simeq 90~{\mathrm ks}$), such effects are negligible with respects 
to the measured fluctuations of the net counts.

\subsection{The CXB flux distribution}

\begin{figure} 
\epsfig{figure=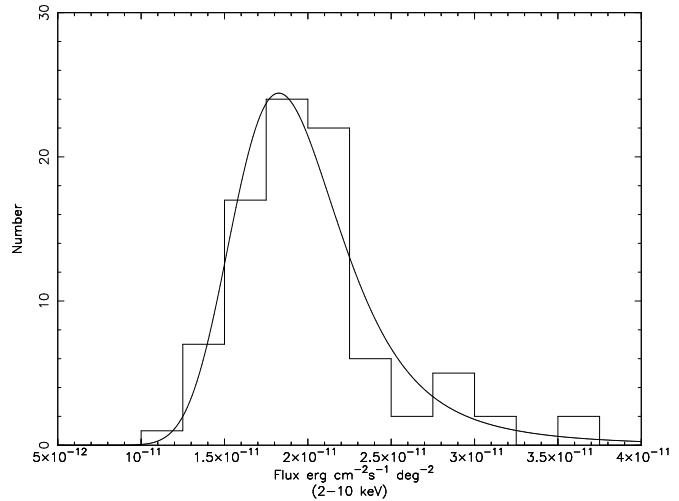,height=9.7cm,width=7.2cm,angle=-90} 
\caption{Distribution of the 2-10 keV Cosmic X-ray Background 
measurements (histogram). The best fit model curve from our ML 
estimation (solid line) is also shown.}
\label{histo} 
\end{figure} 

Net counts between channels 44 and 200 (2.0-9.0 keV) have been 
extracted from the central 8 arcminutes quadrants of the 22 ``blank'' 
fields and converted to fluxes in the 2-10 keV band. 
A power law spectrum with energy index $\alpha = 0.6$ has been 
assumed, as indicated by \asca and \sax deep 2-10 keV X-ray surveys 
(Ueda et al. 1999b; Giommi, Perri \& Fiore 2000). Corrections for the 
Galactic absorbing neutral hydrogen column density ${\mathrm N_H}~$ 
have also been applied. Counts in the channel range 200-220 (9.0-10.0 keV) 
have been excluded since the NXB dominates this energy band. The flux 
distribution of the 88 independent measurements of the Cosmic X-ray 
Background is shown in Figure \ref{histo}. The average flux is 
$\simeq 2.3 \times 10^{-11} \cgs {\mathrm deg}^{-2}$, a value that is 
$\sim 35\%$ higher than that of HEAO1 (Marshall et al. 1980) and slightly 
higher ($\sim 15$\%) than the \asca GIS value (Miyaji et al. 1998) but 
in good agreement with the Wisconsin rocket data (McCammon \& Sanders 1990 
and references therein) and with the MECS measurement obtained from a 
smaller data sample by Vecchi et al. (1999). 

\section{Fluctuation Analysis}

To compute the predicted CXB spatial fluctuations we have adopted an 
analytical approach that, besides the statistical fluctuations due to the 
expected number of X-ray sources, includes counting statistics and the 
effects of MECS vignetting and PSF.

Our method can be summarized as follows. A power law form for the 
log$N$-log$S$ relationship is assumed, 
$N(>S) = K ~ (S/S_0)^{-\gamma}$, 
where $K$ is the number of sources per unit area with flux $S>S_0$.

The expected number $\bar{N}$ of sources in one beam (one MECS quadrant) 
in the flux range from $\bar{S}$ to $\bar{S}+dS$ is then obtained simply 
multiplying the differential of the log$N$-log$S$ with the area of a MECS 
quadrant. Next, the probability of a fluctuation of the number of sources 
is computed assuming Poisson statistics and the corresponding photon counts 
distribution. Counting statistics and vignetting effects are included in the 
procedure by means of two independent probability convolutions with i) the 
Poisson distribution of counts themselves and ii) the off-axis angles 
probability distribution of sources in the MECS FOV. Probability convolutions 
have been computed by means of a Fast Fourier Transform (FFT) technique. 

\begin{figure*} 
\centering{
\epsfig{figure=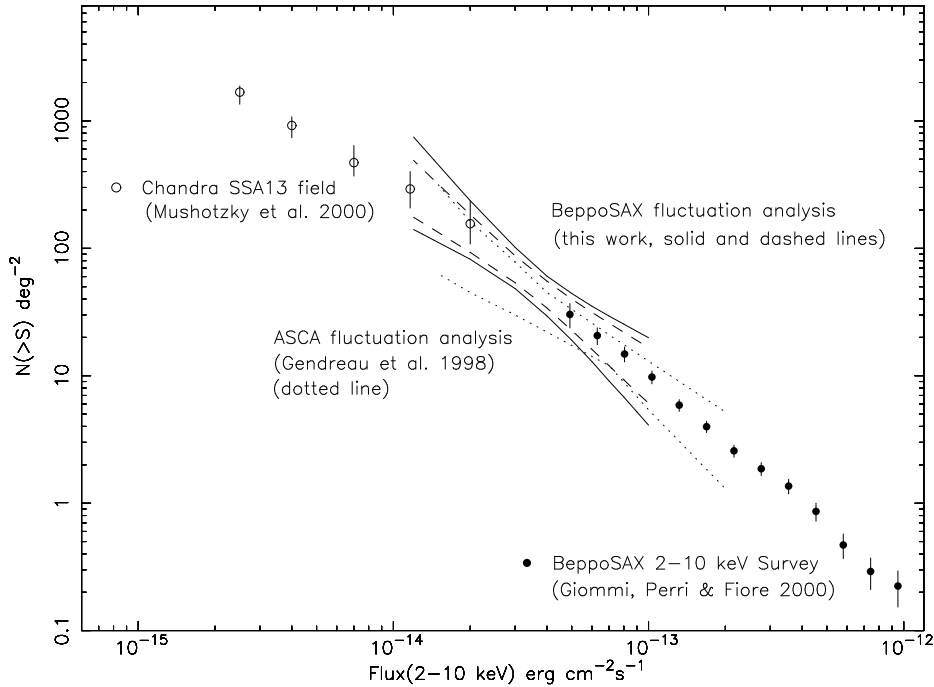,height=14.0cm,width=10.4cm,angle=-90} }
\caption{2-10 keV integral source counts.  Solid and dashed ``bow tie'' 
regions are 90\% and 68\% contours from the present \sax 
fluctuation analysis. Dotted lines are the 1 $\sigma$ contours from 
the \asca fluctuation analysis of Gendreau, Barcons \& Fabian (1998). 
Filled circles represent the \sax 2-10 keV log$N$-log$S$ relationship 
(Giommi, Perri \& Fiore 2000) and open circles are \chandra source 
counts from the SSA13 field (Mushotzky et al. 2000).}
\label{210counts} 
\end{figure*} 

The 2-10 keV MECS PSF size (the radius where 80\% of the photons are collected 
is $\sim 2.6$ arcminutes, Boella et al. 1997b) is non-negligible compared 
to the regions were we extract counts and consequently photons due to sources 
just outside the 8 arcminutes quadrant cannot be neglected. Similarly to the 
\asca fluctuation analysis of Gendreau, Barcons \& Fabian (1998) we have 
then divided a MECS quadrant and ``nearby regions'' with 2 arcminutes size 
boxes and computed the fraction of photons that fall inside a quadrant 
at different FOV positions. 
We have next constructed the off-axis angles probability distribution of sources 
and included this effect with a further convolution. This procedure is 
repeated for all fluxes between $S_{\rm min}$, the flux at which the 
log$N$-log$S$ integrated intensity exceeds the CXB value (as estimated in 
Vecchi et al. 1999), and $S_{\rm max} = 5 \times 10^{-13} \cgs$ (sources with 
$S > S_{\rm max}$ are absent from our data and do not contribute to the 
CXB fluctuations). 
Finally, from the probability convolution of the computed flux distributions 
we obtain the expected distribution of CXB spatial fluctuations corresponding 
to the assumed values of $\gamma$ and $K$.

We have compared the observed CXB spatial fluctuations with a set of model 
curves computed assuming different trial log$N$-log$S$ relationships. 
A maximum likelihood (ML) test has been used to estimate the best fit, 
$68\%$ and $90\%$, ($\Delta S=2.3$ and 4.6) constraints on the 2-10 keV 
log$N$-log$S$ power law slope $\gamma$ and normalization $K$ (as defined by 
equation (2) with $S_0 = 1 \times 10^{-14} \cgs$). The best fit values are 
$\gamma = 1.5 \pm 0.2$ and $K = 336^{+114}_{-58}~{\mathrm deg}^{-2}$ where 
the quoted errors correspond to 90\% confidence level for one interesting 
parameter ($\Delta S=2.7$). The solid line in Figure \ref{histo} shows the 
model best-fit flux distribution which approximates well the observed one 
(a Kolmogorov-Smirnov test gives a probability $>95\%$ that the two 
distributions are equal). 

The 90\% and 68\% constraints of our fluctuation analysis on the 
2-10 keV integral log$N$-log$S$ are shown in Figure \ref{210counts}. For 
comparison the 1 $\sigma$ contours from the \asca SIS fluctuation analysis 
of Gendreau, Barcons \& Fabian (1998) and 2-10 keV source counts from \sax 
(Giommi, Perri \& Fiore 2000) and \chandra (Mushotzky et al. 2000) 
surveys are also plotted.

\section{Simulations}
In order to check our procedure we have carried out extensive simulations of 
MECS images using the data simulator available at the \sax Science Data Center 
(Giommi \& Fiore 1997). This tool fully includes MECS instrumental features 
as telescope vignetting, PSF shape,  non X-ray background and Poisson noise.

We generated 3 sets of one hundred MECS images with pointlike sources 
following 3 different log$N$-log$S$ distributions. Sources have been simulated 
in the flux range $S_{\rm min} - S_{\rm max}$ (see above) with exposure 
time $T = 90~{\mathrm ks}$. 

Each set of simulated images have been analyzed following the same procedure 
used for the real data. The resulting flux distributions have then been 
compared with the model curves and the maximum likelihood test performed. 

We have found that the input values for the power law slope $\gamma$ and the 
normalization $K$ of the log$N$-log$S$ are in all 3 cases within the 68\% 
confidence contours of the fitted ones indicating that no significant bias in 
their estimation affects our analysis.

\section{Summary and Conclusions} 

We have performed a CXB fluctuation analysis based on 22 non-overlapping high 
galactic latitude ($|{\rm b}| > 25\degr$) MECS images of ``blank'' parts of 
the sky. The total exposure time is $\simeq 1.94~{\mathrm Ms}$ while the area 
covered is $\simeq 1.23~{\mathrm deg}^{2}$. The analysis of the CXB spatial 
fluctuations has allowed us to extend the 2-10 keV log$N$-log$S$ relationship 
down to fluxes $\sim 1.5 \times 10^{-14} \cgs$. At this flux level the 
contribution of discrete sources to the 2-10 keV Cosmic X-ray Background is 
$\sim  40-50\%$. The best fit values for the slope $\gamma$ and the 
normalization $K$ of the log$N$-log$S$ are $\gamma = 1.5 \pm 0.2$ and 
$K = 336^{+114}_{-58}~{\mathrm deg}^{-2}$ (quoted errors correspond to 90\% 
confidence level). Figure \ref{210counts} summarizes our results. A good 
agreement with the \asca fluctuation analysis of Gendreau, Barcons \& Fabian 
(1998) and with \sax direct source counts at bright fluxes 
(Giommi, Perri \& Fiore 2000) and a first estimate of the \chandra 2-10 keV 
log$N$-log$S$ (Mushotzky et al. 2000) at faint fluxes is found.

\section*{Acknowledgements} 
\noindent We thank F. Tamburelli for her contribution to the 
development of a new XIMAGE routine and M. Capalbi for her help 
with the \sax archive. We also wish to thank F. Fiore for useful 
discussions and help with the estimation of the MECS Non X-ray 
Background. M. Perri acknowledges financial support from a 
Telespazio research fellowship. Part of the software used in 
this work is based on the  NASA/HEASARC FTOOLS and XANADU packages.

\end{document}